\documentstyle[epsf,twocolumn,pre,aps]{revtex}

\def\beq{\begin{equation}}
\def\eeq{\end{equation}}

\begin{document}
\sloppy

\title{Linear and nonlinear time series analysis \\
       of the black hole candidate Cygnus X-1}
\author{Jens Timmer$^1$, Udo Schwarz$^2$, Henning~U.\ Voss$^2$, Ingo Wardinski$^2$,
Tomaso Belloni$^3$, G\"unther Hasinger$^4$, Michael van der Klis$^3$, and
J\"urgen Kurths$^2$\footnotemark}
\address{$^1$ Zentrum f\"ur Datenanalyse und Modellbildung,
                Universit\"at Freiburg, Eckerstr.~1, D-79104 Freiburg, Germany}
\address{$^2$ Institut f\"ur Physik, Universit\"at Potsdam,
		  D-14469 Potsdam, Potsdam, Germany}
\address{$^3$ Astronomical Institute ``Anton Pannekoek'', University of Amsterdam,
		Kruislaan 403, NL-1098 SJ Amsterdam, The Netherlands}
\address{$^4$ Astrophysikalisches Institut Potsdam, An der Sternwarte 16, 
D-14482 Potsdam, Germany}

\date{\today}
\maketitle

\setcounter{footnote}{1}
\renewcommand{\thefootnote}{\fnsymbol{footnote}}
\footnotetext{Juergen@agnld.Uni-Potsdam.de}
\setcounter{footnote}{1}

\begin {abstract}

We analyze the variability in the X-ray lightcurves of the black hole 
candidate Cygnus X-1 by linear and nonlinear time series
analysis methods.
While a linear model describes the over-all 
second order properties of the observed data well, 
surrogate data analysis reveals a significant 
deviation from linearity. We discuss the relation between 
shot noise models usually applied to analyze these data 
and linear stochastic autoregressive models.
We debate statistical and interpretational issues of surrogate data
testing for the present context.
Finally, we suggest a combination of tools from linear and
nonlinear time series analysis methods as a procedure to test 
the predictions of astrophysical models on observed data.

PACS: {05.40.+j, 02.50.Wp, 97.80.Jp}
\end {abstract}

\section{Introduction}

Cygnus X-1 is one of the best established black hole candidates.
Mass accretion from its primary HDE 226868 leads to 
X-ray emission which exhibits a variability on time scales of tenths
of seconds~\cite{oda71} up to months \cite{kemp1983}. The shorttime 
variability is assumed to be caused by
instabilities of the accretion disk and is usually formally described
by shot noise models~\cite{terrell72,belloni90,lochner91} 
which are a specific kind of point processes. 
These models are inspired by hypotheses about the
physics of the accretion process and the processing of X-rays by 
Comptonization in the neighborhood of the black hole.
Free parameters of these models,
like morphology and distribution of the shots, are usually tuned to fit
the observed energy or power spectra. 

On the other hand, starting from the observed data and characterizing 
the dynamical structure of this observed variability by time series 
analysis methods might yield valuable constraints on astrophysical
models.
This characterization can be, for example, a fit of an explicit model
to the data or the extraction of a feature which captures some
typical structure of the dynamics.
Such a characterization could either inspire new astrophysical models
or could be used for additional tests of the predictions of existing models.
Of course, there is no direct way for a characterization neither
by modeling nor by feature extraction of observed
data to an astrophysical model:
On the one hand, although the goodness-of-fit
of a diagnostic model can be evaluated by statistical tests, these 
tests might have low diagnostic power to detect a misspecification of the
model. On the other hand, a certain feature discovered in the data might be 
generated by many different types of dynamics. 
Therefore, before drawing conclusions about the underlying process
from data analysis, different independent approaches should be
used and the plausibility of a fitted model or an extracted feature 
should be judged in the light of astrophysical knowledge.

The first step of nonlinear time series analysis is usually to study
the structure of a possible underlying attractor.
However, methods from nonlinear dynamics did not succeed to establish a 
low-dimensional attractor for X-ray lightcurves of 
Cygnus X-1~\cite{lochner89}.
It is also important to mention that time series analysis methods usually
assume that the underlying process presents a dynamical system 
in contrast to a shot noise model.

As an alternative to the commonly applied shot noise models,
the linear state space model (LSSM) as a generalization of 
dynamical linear autoregressive models including the observational noise
has been proposed to model
the X-ray variability of active galactic nuclei in~\cite{koenig97}.
Two attractive properties of this approach are, firstly, that the LSSM
can be fitted to the data in the time domain and, secondly, that it 
explicitly takes the observational noise covering the dynamics into account. 
The state space model has been applied to data from Cygnus X-1
in its low state~\cite{pottschmidt98}.
This analysis has revealed that a first order
autoregressive process describes the dynamics of the 
X-ray variability well. This predicts a shot noise model
with an exponential decay and a very specific mode of
excitation of these shots. 

In this contribution, we
analyze X-ray lightcurves of Cygnus X-1 from its low and
intermediate state by the LSSM as well as
by a method which is able to capture deviations from linearity. 
In accordance with~\cite{pottschmidt98}, a scalar LSSM
results in a fit that explains the linear correlations of
the time series well.
However, the nonlinear analysis using a measure
for time reversibility of the process, reveals strong deviations from linearity
on exactly that dynamical time-scale found by the LSSM. 
To interpret this result consistently, we discuss the
mathematical and astrophysical implications of linear stochastic and 
shot noise models.

Finally, we suggest a combination of tools from linear and
nonlinear time series analysis methods as a procedure to test 
the predictions of astrophysical models on observed data.

The organization of the paper is as follows:
In Section~\ref{thedata}  we introduce the data under investigation.
In Section~\ref{methods} we discuss shot noise and linear stochastic
models and their relation. Furthermore, we explain how we use the
method of surrogate data to test for time reversibility.
Section~\ref{results} presents the results, which are discussed in
Section~\ref{discussion}.

\section{The data}
\label{thedata}

The data were recorded with the Proportional Counter Array (PCA)
on board the Rossi X-ray Timing Explorer (RXTE).
The X-ray activity of  Cygnus X-1 is classified as low, intermediate,
and high, depending on the mean count rate~\cite{belloni96}.
Our analysis is based on two data sets:
The first data set was recorded on 22nd May 1996, 19:05:12 - 19:48:02,
while Cygnus X-1 was in its intermediate state~\cite{belloni96}.
The energy range was 2.0 - 14.1 keV (channel range: 0-35).
The sampling frequency was 256 Hz and the data set consists of 
655,360 data points. The mean number of counts per bin 
was 38.3 with standard deviation 10.0. 
The second data set was recorded on 12th February 1996, 
9:37:20 - 10:03:06, while Cygnus X-1 was in its low state. The energy 
range was 2.0 - 9.9 keV (channel range: 0-35). The sampling frequency 
was 256 Hz and the data set consists of 394,752 data points.
The mean number of counts per bin was 18.7 with standard deviation 7.1. 
Figure~\ref{fig_data} displays a 3~s segment of the first data set.
A part of the variability of the data is explained by the fact
that the recording process is a counting process. This leads
to additive uncorrelated observational noise which is Poisson
distributed. Due to the high mean count rate this Poisson
noise is well approximated by Gaussian noise.

\section{Methods} \label{methods}

\subsection {Shot noise processes} 
\label{shot_noise_processes}

Shot noise processes are a specific type of point processes~\cite{cox80}.
Point processes are characterized by a 
probabilistic law that some event happens at a certain time.
For the simplest form of a shot noise model the
probabilistic law of occurrence of
events follows a Poisson process and the event is an exponential
decay with initial value $M$ and decay time $\tau$. 
A Poisson process is defined by the property that the probability 
of an event to take place in a time interval $(t,t+\Delta t)$ 
is proportional to $\Delta t$ in the limit of small intervals:
\beq \label{poisson}
  \lim_{\Delta t \rightarrow 0} \mbox{prob (Event in} (t,t+\Delta t)) = 
		\rho \Delta t \; ,
\eeq
where $\rho$ denotes the intensity of the process. The sampled time 
series consists of a superposition of the single shots at times $T_j$
whose occurrence follows Eq.~(\ref{poisson}), i.e.,
\beq \label{shot_noise}
  x(t_i) =\sum_j M \, \Theta (t_i-T_j) \, e^{-(t_i-T_j)/\tau}
\eeq
with $\Theta (z)=1$ if $z\ge 0$,  $\Theta (z)=0$ if $z<0$.
We call this process the classical shot noise process.

The power spectrum of this process (\ref{shot_noise}) is given
by~\cite{papoulis}:
\beq \label{shot_noise_spec}
 S(\omega) = \frac{M^2 \, \rho}{1/\tau^2 + \omega^2}, \quad \omega \ne 0.
\eeq

The classical shot noise has  already been proposed in Ref.~\cite{terrell72}
to describe the observed variability of the lightcurves of Cygnus X-1.
It consists of exponentially decaying shots with fixed initial value 
which occur in time with a constant rate of probability. 
Several generalizations
have been proposed:
Shots with a decay rate drawn from
a certain distribution have been suggested in
\cite{belloni90,nolan81,miyamoto89}.
A distribution for the initial values of the shots
was considered in~\cite{lehto89}. Vikhlinin et al.~\cite{vikhlinin94} 
introduced interactions
between different shots. Furthermore, the simple exponential form
was replaced by more complicated time courses showing
an initial increase from zero to a maximum value followed
by a decay to zero~\cite{pottschmidt98}. These types
of profiles are supported by Monte Carlo simulations 
of astrophysical models of the X-ray processing by spatially
resolved Comptonization in a cloud of hot electrons
surrounding the accretion disk~\cite{dove97}.

For some generalized shot noise models
the power spectra can be calculated
analytically
\cite{lochner91,papoulis};
otherwise they have to be estimated from simulated data.

\subsection{Linear stochastic dynamical systems} 
\label{methods_linear}

In contrast to shot noise processes given by 
Eqs.~(\ref{poisson},\ref{shot_noise}),
continuous dynamical systems are given by a differential equation
\beq \label{dyn_sys}
  \dot{\vec{x}} = \vec{f}(\vec{x},\vec{\epsilon}) \; ,
\eeq
where $\vec{\epsilon}$ denotes random perturbations which
might influence the time evolution of the dynamics. An attractive feature of 
modeling time series by dynamical systems is that
the specific form of $\vec{f}(\vec{x},\vec{\epsilon})$
might provide insight in the physics at work, see 
\cite{voss99,hegger98} for two examples from physics and
\cite{timmer98d,michalek99c} for application to physiological time series.

In the simplest case if $f(.)$ is linear in $\vec{x}$ and the
dynamical noise $\vec{\epsilon}$ is Gaussian distributed and additive, 
the system represents
linear combinations of damped oscillators and relaxators that
are driven by Gaussian noise. Since the model is linear, 
all information about the model is captured by the power spectrum.
For a scalar dynamics:
\beq \label {ar_cont_proc}
   \dot{x} = - \alpha x + \epsilon, \quad \epsilon \sim WN (0,\sigma^2) \;,
\eeq
 the spectrum is given by:
\beq \label {ar_cont_spec}
  S(\omega) = \frac{\sigma^2}{ \alpha^2 +  \omega^2} \; .
\eeq
It is important to emphasize that first order linear stochastic dynamical 
systems have the
same $\omega$-dependence of the spectrum as the classical shot noise model, 
see Eq.~(\ref{shot_noise_spec}).

Most often, $\vec{x}$ cannot be observed directly, but 
only a scalar function $g(\vec{x})$.
Furthermore, the observation
$y$ might contain additive measurement noise, denoted by $\eta$:
\beq \label{obs_eq}
 y=g(\vec{x}) + \eta \; .
\eeq
While the noise $\vec{\epsilon}\/$ in Eq.~(\ref{dyn_sys}) drives the dynamics,
the measurement noise $\eta\/$ in Eq.~(\ref{obs_eq})
only disturbs the observation of the system.
For the case of a linear dynamical system, Eq.~(\ref{ar_cont_proc}),
with white additive observational noise of variance $R$,
the spectrum reads:
\beq  \label{lssm_spec}
  S(\omega) = \frac{\sigma^2}{ \alpha^2 +  \omega^2} + R \; .
\eeq
Since measured data are sampled, discrete time dynamical models
\beq
  \vec{x}(t) = \vec{h}(\vec{x}(t-\Delta t),\vec{\epsilon}(t)) \; ,
\eeq
are often used. 
If both the dynamical and the measurement noise are Gaussian distributed,
and the functions $\vec{h}$ and $g$ are linear, i.e.,
\beq
\label{lssm}
\begin{array} {llll}
    \vec{x}(t) & = & A \vec{x}(t-\Delta t) +  \vec{\epsilon} (t) ,
	&	\quad \vec{\epsilon} (t) \sim { N} (0,Q) \; , \\
	y(t) & = & C  \vec{x}(t) + \eta (t),
		& \quad \eta (t) \sim { N} (0,R) \; ,
\end{array}
\eeq
the  linear state space model (LSSM) as a generalization
of the well known autoregressive (AR) models results.
They represent discrete time versions of the continuous time
linear stochastic models.
The matrix $A$ determines the dynamics of the unobserved state
vector $\vec{x}(t)$. Its dimension reflects the order of the process.
The vector $C$ maps the state vector to the observation.
In the case of a scalar dynamics, $A$ is related to 
the relaxation time scale $\tau$ by $\tau=-1/\log|A|$.
The mathematical formalism of the LSSM and procedures to estimate its parameters
are described in detail in~\cite{timmer98d,honerkamp_buch}.

To test the consistency of a fitted model with the data, 
at least three criteria should be applied.\\
1. The variance of the prediction residuals does not decrease 
significantly for larger model dimensions. \\
2. The spectra calculated from the fitted LSSM for larger model dimensions
coincide.\\
3. An appropriate model should turn the correlations in the data
into prediction residuals consistent with white noise. 
In the frequency domain this hypothesis can be tested by
comparing the periodogram of the residuals with the expected 
straight line in the case of white noise by the Kolmogorov-Smirnov test 
\cite{recipes}.

\subsection{Noise reduction}

Measured time series of natural systems often contain a large amount 
of additive observational
noise. The fitted LSSM can be applied as a linear filter to perform 
a noise reduction on the data even if it is misspecified as a 
dynamical model of the underlying process.
If the LSSM describes the second order properties of the process
correctly, the LSSM is the optimal linear filter~\cite{honerkamp_buch}.

Algorithmically the noise reduction is achieved by first applying
the Kalman filter, which yields an estimate of $\vec{x}(t)$ based
on the observed data $y(1), y(2), \ldots, y(t)$. Then the so-called
smoothing filter is applied backwards in time to obtain estimates 
$\hat{\vec{x}}(t)$ based on the whole data set~\cite{honerkamp_buch}. 
The possibility to apply
this smoothing filter relies on the property of linear stochastic
processes to be time reversible, see Section~\ref{vergleich}.
Multiplication of $\vec{x}(t)$ by the estimated $C$ yields an estimate of
the noise-free  scalar observable $y(t)$.

The statistical properties of the estimated $\hat{y}(t)$ can be understood
in the frame of Bayesian estimation, see~\cite{kitagawa96} for a detailed
discussion. The model with its fitted parameters
represents a prior on the smoothness of the hidden $\vec{x}(t)$. 
Conditioned on this prior a maximum likelihood estimate 
of $y(t)$ is obtained. The estimated time series is the most probable
one assuming the validity of the model, Eq.~(\ref{lssm}). 

It should be emphasized
that the estimated time series does not represent a typical
realization of the fitted model used as prior. Even if the fitted
model is the true one, the estimated time course is a slightly
low-pass filtered version of a typical realization. If the 
fitted model is, however, not the true model, the estimated time series
will show statistical properties which, literally spoken, lie
between those of the process which generated the data and
the model used as prior. Especially, if the true process
is nonlinear showing a strong time irreversibility, this quantity might 
be reduced for the estimated time series. Thus, the procedure
does not lead to false positive results.

\subsection {The relation between linear models and shot noise models}
\label{vergleich}

Linear autoregressive and shot noise processes are both stochastic processes. 
The randomness driving these processes usually
reflects the restricted knowledge
about the dynamics at work. Often, the dynamics is exposed to
numerous influences that  cannot be taken into account explicitly. 
Even if these influences are deterministic in nature they effectively act as 
random influences due to their large number.
The characteristic difference between autoregressive and shot noise processes
is the way the randomness  enters the process: i) In dynamical
 processes it describes a random force that influences the dynamics 
in every instant of time. ii) In point processes
it acts as a trigger that generates a certain event only at certain
points in time. 

However, there is a formal connection between the classical
shot noise process and the scalar linear stochastic dynamical process. 
Formally, and ``not in the spirit of point processes''~\cite{cox80},
one can transform Eq.~(\ref{shot_noise}) into 
\beq
  x(t) = (1-\Delta t/\tau) \, x(t-\Delta t) + \epsilon(t) \; ,
\eeq
where $\epsilon(t)$ has the specific form:
\beq \label{epsilon}
  \epsilon(t) = \left\{\begin{array}{cl} 
		0 & \mbox{with probability } 1-\rho \Delta t\\
 		M & \mbox{with probability } \rho \Delta t
	\end{array}\right. \; .
\eeq
Thus, for $\rho \Delta t \approx 1$ and $M$
following a Gaussian distribution,
there is a formal equivalence between the scalar linear autoregressive process
and the classical shot noise process which is characterized by its
exponentially decaying shot profile.
In practice $\Delta t$ corresponds to the sampling interval. 
The condition  $\rho \Delta t \approx 1$ means that the process is highly
undersampled, since single shots are not resolved. The required Gaussianity
of the distribution of the initial values of the shots does not meet the
physical constraint of positivity in the astrophysical context of X-ray bursts.
In the limit  $\rho \Delta t \approx 1$ it might be an effective
description resulting from the superposition of the unresolved Poisson
process.

In summary, scalar linear dynamical processes are a certain formal limiting 
case of  shot noise models.
Only in the case of linearity, there is no interaction between 
the excitations and time course of the shots.
It should be noted that, in general, nonlinear stochastic dynamical systems 
cannot be formulated as a formal limit of shot noise models.

\subsection{Beyond linear models: Time irreversibility} 
\label{behind}

An important property of linear Gaussian
processes is time reversibility, i.e., the statistical properties
of the process are the same forward and backward in time~\cite{weiss75}. 
An intuitive
explanation is that the statistical properties of these processes are
completely captured by the autocorrelation function, which is by
definition symmetric under time reversal. Shot noise processes 
with non-symmetric shot profiles are not time reversible as are 
many nonlinear dynamical systems.
The Gaussianity of the noise $\epsilon(t)$ of a linear 
autoregressive process
is crucial for the time reversibility. Any deviation from Gaussianity
leads to time irreversibility even in the case 
of linear dynamics~\cite{weiss75}.
This is of special interest in view of Eq.~(\ref{epsilon}).
While time reversibility
has been used to test for nonlinearity in dynamical systems,
\cite{theiler92,timmer93,vanderheyden96,voss98},
we will use it here as an indicator for a shot noise model.
A test for time irreversibility in this context
 will be discussed in the next section.

\subsection{Nonlinear analysis: The method of surrogate data}
\label{methods_nonlinear}

The theory of nonlinear dynamical systems offers notions
to characterize processes beyond linearity, see
\cite{abarbanel96,kantz97} for a review.
Different quantities have been invented to reveal whether
an observed time series is a realization of a chaotic system;
among others, the correlation dimension~\cite{grassberger83a},
Lyapunov exponents~\cite{wolf85},
and nonlinear forecasting errors~\cite{sugihara90}.
It has been observed later that due to the finite size of data,
noise, and linear 
correlations, the algorithms to calculate these quantities
can give false positive results.

To test the reliability of the results,
the method of surrogate data has been invented independently by different
authors, e.g.~\cite{elgar84,grassberger86,Kurths87,kaplan90,kennel92}, 
but has been made most popular by~\cite{theiler92}.
It has found wide applications in the analysis of astrophysical 
\cite{Kurths87,Provenzale94,Vossp1,leighly97}, geophysical 
\cite{Vossp9,Smith94a,Zoeller98} and biophysical
\cite{schiff92,schiff94a,schiff94b} data.

The general idea is to simulate time series whose statistical properties 
are constrained to the null hypothesis one wants to test 
for~\cite{schreiber98}.
In testing for linearity this is achieved by randomizing the phases
of the Fourier transform
of the data and transforming the result back to the time domain. 
A possible static nonlinearity in the observation, $g(\vec{x})$ in
Eq.~(\ref{obs_eq}),
is known to produce spurious significant results~\cite{rapp94}.
Therefore, a
proper adjustment of the distribution of the time series data is performed.
For many realizations of time series from this procedure, 
the same algorithm as to the
original data is applied leading to a distribution of the feature
calculated by the algorithm assuming linearity.
A significant difference between
the distribution of the feature produced by the algorithm for the surrogate 
data and the original data is taken as an indication that
the process underlying the original is not a Gaussian, stationary,
stochastic, linear one. A significant result of the
test does not necessarily indicate chaoticity of the process,
since this is only one possibility to violate the null hypothesis.

Former analysis revealed that it is unlikely that the Cygnus X-1
as well as other comparable X-ray sources represent
a low-dimensional chaotic system~\cite{lochner89,norris89,lehto93}. 
Therefore, we apply 
the surrogate data test to look for deviations from the null
hypothesis in general.

The results of the surrogate data test for a feature $f$ are usually reported 
as significance $S$ :
\beq \label{significances}
  S = \frac { |f-{\left<f\right>}_{surr}| }{\sigma_{surr}}\;,
\eeq
where ${\left<f\right>}_{surr} $ denotes the mean of the distribution of 
the feature for the surrogates and $ \sigma_{surr} $ its standard deviation.
Assuming a Gaussian distribution for the feature a value of $S=2.6$
corresponds to a significance level of $\alpha=0.01$. 

We propose here a surrogate data analysis based on 
time reversibility.
Generalizing a suggestion of Weiss~\cite{weiss75}, a simple measure denoted by
$Q(m)$ for a deviation from reversibility for a certain
time lag $m$ was introduced in~\cite{theiler92}:
\beq \label {q_statistic}
  Q(m) = \frac{ \left<(x(t+m)-x(t))^3\right>}{\left<(x(t+m)-x(t))^2\right>} \;.
\eeq
More complex measures for time irreversibility based on
conditional, respectively joint probability distributions
are described in~\cite{timmer93,vanderheyden96,voss98}.

Since it is not clear beforehand at which lag $m$ a possible
deviation from the null hypothesis might result in a significant 
$Q(m)$~statistics, the significances $S(m)$ will be evaluated for 
all lags up to a maximum lag.
This leads to the statistical problem
of multiple testing. It is important to emphasize that this has an impact on 
the level of significance $\alpha$, i.e., the probability
to reject the null hypothesis although it is true. 
If the null hypothesis is tested in $n$ independent tests at the level 
 $\alpha$, the
probability to reject the null hypothesis at least once is given
by
\beq \label{bonfe1}
  \tilde{\alpha} = 1 - (1 - \alpha)^n \; .
\eeq
For example, for $\alpha = 0.01$ and $n=10$, the actual significance
level $ \tilde{\alpha} $ is $0.1$, leading to a ten times higher
probability for an incorrect rejection of the null hypothesis
than expected. A simple cure to this problem is the
Bonferroni-correction~\cite{sachs84}. Therefore, Eq.~(\ref{bonfe1}) 
is solved for $\alpha$:
\beq \label{bonfe2}
  \alpha = 1 - (1 - \tilde{\alpha})^{1/n} \; .
\eeq
Since $\tilde{\alpha} \ll 1$, the right hand side
of Eq.~(\ref{bonfe2}) can be approximated in first order, resulting in
the simple rule:
\beq \label{bonfe3}
  \alpha = \tilde{\alpha}/n \; .
\eeq
This procedure is known to be extremely conservative, i.e., while
it guarantees that the significance level is correct, the test
loses its diagnostic power to detect a violation of the null hypothesis.
For some test statistics, procedures are 
known to obtain tests that have the correct significance level as well 
as a good diagnostic power, see e.g.~\cite{sachs84,miller82,lausen92}. 
It is not known to the authors, how to apply an analogous strategy to the 
$Q(m)$~statistics. The main problem is
that the correlations in the time series produced by the underlying dynamics
of the process lead to correlations between the $Q(m)$~statistics
for different lags. Thus, the only cure known to the authors 
is to check whether the results of 
an analysis of one time series can be reproduced by the analysis 
of independent measurements.  Therefore, we subdivide our time series
into segments of length 20,000 data points each and calculate the averaged
$Q(m)$~statistics and its confidence interval.

To reveal the expected behavior of the $Q(m)$~statistics for shot noise
processes, we simulate an exponential shot noise process with
intensity $\rho=0.1$, $\tau=15$, initial values $M_i$ drawn from a uniform 
distribution in the interval [0,1], and apply the $Q(m)$~statistics.
Figure~\ref{fig_q_true}a shows a segment of the simulated data.
Figure~\ref{fig_q_true}b and c display the $Q(m)$~statistics
and the significances $S(m)$ 
for different lags $m$ based on a realization of the process of length 
20,000 data points. The monotonically decaying behavior
of the $S(m)$ curve does not depend on the intensity, the relaxation time
or the distribution of the shot noise process. Of course, the
quantitative behavior does. 
Classical shot noise and first order linear stochastic dynamical
systems can not be discriminated by linear methods since their
spectra coincide. The simulation shows that higher order
statistical properties allow for a discrimination.
Next we apply this concept to the analysis of measured data.

\section {Results} \label{results}

We discuss the results for the time series of the intermediate state
in detail.
For the linear analysis, the results for the
intermediate and low state data  are comparable.
Differences for the
nonlinear analysis will be presented in more detail in 
Section~\ref{nonlinear_analysis}.

\subsection{Linear analysis by state space models}

We fit linear state space models (LSSM), Eq.~(\ref{lssm}), of increasing 
dimension to segments of the intermediate state time series of length 20,000.
In accordance with the results of~\cite{pottschmidt98} for 
the low state, the residual variance is
constant for all models of dimension larger than zero.
Furthermore, the analysis reveals an equal contribution of signal 
and noise to the total variance of the time series.

Figure~\ref{fig_spec_lssm} displays the periodogram of the first segment
and the spectra calculated from fitted one- to three-dimensional
models on a log-linear and on a log-log scale. 
The spectrum calculated from the fitted parameters well
explains the over-all periodogram of the data. Furthermore,
there is no significant difference between the spectra
of fitted different dimensional processes.
The relaxation time of the scalar model is 14.2 sampling units
corresponding to 55 ms. 
The Kolmogorov-Smirnov test does not reject the hypothesis
of white noise residuals at the 1\% level of confidence.

With respect to the dimension of the model, a fit of LSSMs of dimension 
one to three to the remaining 31 segments confirms
the result for the first segment. For the pieces of 20,000
data points as well as for the whole data set the spectra
calculated from the estimated parameters
do not differ from the spectra of the scalar model. 
The estimated relaxation
times range from 12.4 to 17.4 sampling units, corresponding
to 48 to 68 ms. For the data set from low state the qualitative
results of the linear analysis are the same as for the intermediate
state, but the relaxation times range from 40 to 56 sampling units,
corresponding to 150 to 220 ms, in accordance with the results
reported in Ref.~\cite{pottschmidt98}.

Linear analysis methods, like spectral analysis, only capture the second 
order statistical properties of a process. For linear processes the higher 
order properties are a function of the second order correlations.
This does not hold for nonlinear processes. Therefore,
it could be possible, that there is some nonlinear dynamics at work
in the process under investigation which is invisible for a linear analysis. 
If such nonlinear dynamics can be described by Eq.~(\ref{dyn_sys}),
it can be concluded that its dimension is not larger than one. Any higher 
dimensional continuous-time system would have led to a difference between
the spectra of the one and the higher dimensional LSSMs, since it would 
produce linear correlations for an order of at least the dimension 
of the process.
In the same line of argument, a nonlinear first order dynamical
process should have effected the higher order spectra.
Thus, the linear analysis strongly suggests a linear stochastic
first order process for a description of the data in the frame of
dynamical systems.

\subsection{Nonlinear analysis} 
\label{nonlinear_analysis}

First, we apply the surrogate data based search for deviations
from linearity as described in Section~\ref{methods_nonlinear}
to segments of length 20,000 up to a maximum lag
of 1000 sampling units corresponding to 3.9 s of the observation.
We use 100 surrogate data sets to estimate the mean and
the variance of the $Q(m)$~statistics, Eq.~(\ref{q_statistic})
 for the null hypothesis
of linearity to calculate the significances $S(m)$, Eq.~(\ref{significances}).

For the first segment,
at above lag 800 the significance $S(m)$ of the $Q(m)$~statistics  for
time reversibility results
in a value larger than 4 (Figure~\ref{fig_q_roh}a).
This corresponds to a 
probability for the null hypothesis smaller than $10^{-4}$.
As discussed in Section~\ref{methods_nonlinear}, the results
of the nonlinear analysis by the surrogate data method
using the $Q(m)$~statistics has to be based on the
consistency of the results for independent measurements due to the
multiple testing problem.
Figures~\ref{fig_q_roh}b--d display the results for the following 20,000
data point segments of the time series.
There is no consistent deviation from the null hypothesis
for any lag. 

Linear analysis reveals that the signal to noise ratio
is equal to one if measured in relative amplitudes.
This large amount of observational noise diminishes
the diagnostic power of the surrogate data test to detect a possible
time irreversibility. 
As discussed in Section~\ref{methods_linear},
the LSSM can be applied to estimate the noise-free dynamical time
series within a Bayesian framework.
Figure~\ref{fig_q_smooth1} displays the
results for the Kalman (and smoothing)-filtered
data based on the one-dimensional LSSM analogous to Fig.~\ref{fig_q_roh}. 
For large lags no significant changes appear apart from a smoother
behavior of the curve which results from the low-pass filter property
of the estimation procedure as discussed on Section~\ref{methods_linear}.
But for small lags the behavior of the curves changes:
Figure~\ref{fig_q_smooth2} shows the significances $S(m)$ of the 
$Q(m)$~statistics for the first 100 lags.
Consistently, a significant deviation from linearity is found for
exactly those lags up to the time scale of approximately 15 sampling units 
that was found as typical time scale by the linear analysis. Note 
that the resulting
$S(m)$ curves for the Kalman-filtered data resemble the decaying curve
expected for a shot noise model, Fig.~\ref{fig_q_true},
 while the raw data suggest a maximum at around 10 sampling units.
The similarity of the results for larger time scales and the
differences for short time scales can be interpreted in the
frame of shot noise models. For lags much larger than the 
relaxation time of the shots, the data are independent
and the $Q(m)$~statistics is expected to vanish.
The appearance
of the $S(m)$ is determined by correlated fluctuations,
as discussed in Section~\ref{methods_nonlinear}.
For time scales smaller than the relaxation time of the
shots, the $Q(m)$~statistics is significantly different from
zero, see Fig.~\ref{fig_q_true}. The difference between
the results for the raw and the Kalman-filtered data
is an effect of the lag dependent signal to noise ratio.
This is most pronounced for the shortest lags,
since the time-course of each shot is continuous,
but the observational noise is discontinuous, leading to
a decreasing signal to noise ratio for smaller lags.
This is the reason why $S(m)$ tends to zero for lags
close to zero for the raw data.

Since the Kalman filter is linear, it is not expected to
lead to artificial results. This has been confirmed
in a simulation study.
We use the fitted one-dimensional LSSM to generate
data and calculated the significance $S(m)$ of the $Q(m)$~statistics 
for these data and data obtained by the Kalman filter.
The results are displayed in
Fig.~\ref{fig_artis} and show that the Kalman filter does not
produce spurious results for processes that are time reversible.
Simulation studies using shot noise processes with added
observational noise show that the Kalman-filtered data
reproduce the behavior of the $S(m)$ curve for shot noise processes
as displayed in the Fig.~\ref{fig_q_true}.
Thus, the significant results are
not due to the Bayesian estimation by the Kalman filter
(Section~\ref{methods_linear}).
This is reasonable since in the worst case this linear filtering 
``pulls'' the data in the direction of behaving more linear.
That means that an existing time irreversibility would be decreased,
but no spurious significant effects are introduced.

Figure~\ref{fig_q_all} shows the mean and 2$\sigma$ confidence region of
the significance $S(m)$ of the $Q(m)$~statistics obtained from the 32 
segments of length 20,000 based on the raw and the noise-reduced
time series from the intermediate state.
Figure~\ref{fig_q_all_low} displays the
corresponding plot for the 19 segments from the low state time series.
For both data sets the $S(m)$ curves for the raw and the Kalman-filtered
time series are statistically indistinguishable
for larger lags. Significant differences arise only for small lags.
Based on the analysis of the raw data, any kind of shot noise model
would be rejected. 
For the analysis based on the Kalman-filtered data, the $S(m)$ curve for 
the low state time series suggests a classical shot noise model 
by its decay for small lags and insignificant values for larger lags,
compare Fig.~\ref{fig_q_true}. For the intermediate state time series,
a significant maximum occurs at a lag $m$ of 30 sampling units, corresponding
to 117 ms. This maximum cannot be reproduced by a simple
shot noise model and calls for more complex processes discussed in 
Section \ref{shot_noise_processes}.

For both time series, our analysis shows that the linear state space model 
is not an appropriate model to describe the data, since the 
significant time reversibilities calculated based on the
fitted models contradict the assumption of these models.
It is, however, important to note that the LSSM can be used to perform
an efficient noise reduction.

\section {Discussion} \label{discussion}

We have developed methods and have discussed how it is possible to 
decide based on measured data whether a time series that even comprises a 
large amount of additive observational
noise has been produced by a scalar linear stochastic 
dynamical system or a shot noise process. 
We have shown that linear spectral analysis does not allow for a discrimination.
The nonlinear property of time irreversibility of shot noise processes
form the basis for a significant distinction. A straightforward evaluation
of this feature is hampered by the statistical problem of multiple
testing and effects of additive observational noise. We have discussed
how these problems can be overcome.

We have applied methods from linear and nonlinear time series
analysis to two X-ray variability lightcurves of the black hole candidate
Cygnus X-1. The first time series was recorded while Cygnus X-1 was in
an intermediate state~\cite{belloni96}, the second represents the low
state. Such data are usually described by shot noise models,
a specific kind of point processes. 
Although  point processes are fundamentally different from 
dynamical systems, they share some properties
with the latter. First, the spectrum of the classical shot noise
process coincides with that of a scalar continuous time linear
Gaussian stochastic process. Second, most shot noise models
share the property of most nonlinear dynamical systems of being
time irreversible.

Firstly, we have fitted linear state space models
(LSSM) of increasing dimension to segments of the data.
The variance of the prediction residuals is not decreasing for models of
dimension larger than zero and the spectra calculated from
the fitted parameters of the different models coincide,
suggesting a scalar dynamical model.
Testing the consistency of the prediction residuals with white noise
has revealed a good over-all fit.
The linear analysis shows that if the process is a
dynamical system, it is linear and one-dimensional. Any higher dimensional
or continuous-time nonlinear dynamical systems would have led to differences 
between one and higher dimensional LSSMs with respect to the spectra
calculated from the fitted parameters and the variance of
the prediction residuals.
Furthermore, the analysis suggests a signal to noise ratio of one.

Fitting a LSSM to data in the time domain is asymptotically
equivalent to fitting its spectrum to the periodogram 
of the data in the frequency domain~\cite{dahlhaus84}.
The spectrum of the classical shot noise process
is identical with the spectrum of a first order linear dynamical process.
Thus, even if a goodness-of-fit test in the 
frequency domain does not reject a LSSM, no discriminating 
conclusions can be drawn with respect to the question
whether a dynamical system or a shot noise process has generated the data.
Therefore, astrophysical interpretations of the parameters
of fitted LSSMs~\cite{pottschmidt98,koenig97a,koenig97b}
should be treated with care.

Astrophysical studies indicate that the
processes under investigation follow some kind of shot noise model 
\cite{terrell72,belloni90,lochner91,pottschmidt98,miyamoto89,dove97,negoro94,liang84,wilms97}.
In general, shot noise models are not reversible in time.
Surrogate data testing for time irreversibility for different lags introduces
the multiple testing problem. 
Therefore, we have investigated whether consistent results could be obtained
from an analysis of segments of the time series.

For the raw data of the low state time series,
no significant deviation from linearity has been detected.
However, we have found a double
well behavior of the $Q(m)$~statistics in the case of the
intermediate state data (Fig.~\ref{fig_q_all}).
Both results contradict a simple shot noise model.
This might have been caused by the low signal to noise ratio.
In the frame of Bayesian estimation based on a fitted LSSM,
we have applied the Kalman filter to get a noise-reduced time series.
Based on these noise-reduced data, we have found a significant deviation from 
linearity at that time scale found by linear analysis
that are in accordance with results for simulated data from 
a simple shot noise model.
While the results for the low state time series are
in agreement with a simple shot noise model with independently decaying
shots, the intermediate state time series shows a more
complex behavior. Apart from the decay for small lags
the significances show an additional distinct maximum.
Our results are based on the estimated noise-reduced time series
obtained by the LSSM.
Any noise reduction procedure imposes assumptions about
the underlying process and might lead to artifacts if the assumptions
are not met as in the present study. In the case considered here  
a violation of the assumptions of the model, in the worst case, leads to 
less significant results since the filter is linear. 
Thus, the procedure is statistically
conservative even if the model is misspecified.

By its qualitative difference to the results for simple shot noise
models for the intermediate state time series, the $Q(m)$ statistics as 
a measure for time irreversibility
poses a constraint on astrophysical models for this phenomenon.
It has been shown that the classical shot noise model 
(\ref{poisson}--\ref{shot_noise_spec}) does not satisfactory describe
the process under consideration~\cite{belloni96}. 
Therefore, one has to search for more complex models.
For such models the significance of the
$Q(m)$~statistics (Fig.~\ref{fig_q_all}) provides an
additional and independent test beyond the usually applied energy and power
spectra. For example, our results exclude shot noise models
with symmetrical rise and decay of the shots as discussed in  
\cite{negoro94}, since such models would not lead to a violation
of time reversibility. In general, one has to Kalman-filter the data 
generated by the proposed model in the same way as the observed 
data and test the compatibility of the resulting $S(m)$ curve
statistically.

No explicit test to decide whether a dynamical system or a shot noise 
process underlies a measured time series is known to the authors.
Summarizing the results from the linear and the nonlinear time series
methods, the analysis strongly suggests that a shot noise model is at work. 
This is in accordance with astrophysical considerations:
X-rays undergo multiple Compton scattering in the corona of hot electrons
surrounding Cygnus X-1. The shots represent the projection
of this spatio-temporal, reaction-diffusion like processes on the time axis.
The loss of spatial resolution causes that the resulting process
cannot be formulated as a dynamical system anymore.
This reveals an interesting aspect of surrogate data testing
that might also apply for other applications~\cite{leighly97}.
Initially, testing by surrogates was introduced to support the 
detection of chaotic dynamics. Later, it was recognized that
a rejection of the null hypothesis of linear, stochastic, 
stationary, Gaussian dynamics does not necessarily indicate
chaos, i.e., a special type of nonlinear, stationary, deterministic
dynamics, since there are other possibilities to violate
the assumptions of the above null hypothesis
\cite{kaplan95,barnett98,Aschwanden98,timmer98e}.
Furthermore, surrogate data testing was characterized as
not too informative if simple inspection of the data
reveals a deviation from the null hypothesis~\cite{kaplan95}.
In the present case, the linear analysis looks promising at first 
sight rendering the surrogate data test informative.  
But here, the reason for a significant surrogate data test is not
chaotic nonlinearity, but the projection from the spatio-temporal 
into the temporal domain. Thus, the X-ray variability
data offer a new possibility for a rejection of the null hypothesis
of a linear dynamical system: The system is not a 
dynamical system of the form 
$\dot{\vec{x}} = \vec{f}(\vec{x},\vec{\epsilon})$ at all.

In summary, following a quotation of G.E.P.~Box: ``All models are
wrong, but some are useful'', we propose the use of the
misspecified linear state space model together with the measure of 
time reversibility inspired by nonlinear dynamics as an additional test to
the usually applied energy and power spectra to evaluate the validity 
of astrophysical shot noise models on measured data.

\section {Acknowledgments}

The authors acknowledge the stimulating atmosphere of
the Isaac Newton Institute where parts of this article
were written.
J.T.~acknowledges the hospitality of the University of Potsdam.
H.U.V.~acknowledges financial support by the Max-Planck-Gesellschaft.



\section*{Figure captions}

\begin{figure}
\parbox{\textwidth}{
\epsfxsize=0.6\textwidth \epsfbox{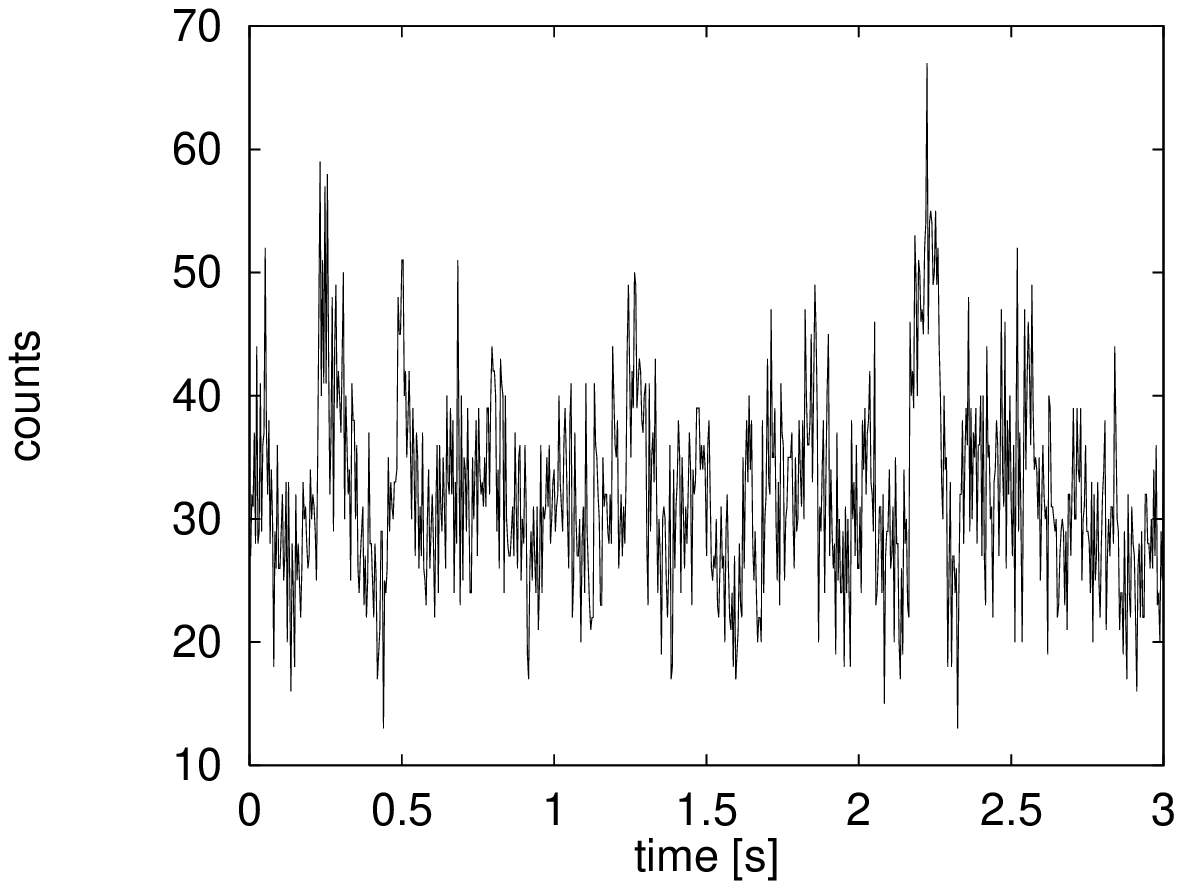}}
\caption[]{\label{fig_data}}
\end{figure}

\begin{figure}
\begin{center}
\parbox{\textwidth}{
\epsfxsize=0.5\textwidth \epsfbox{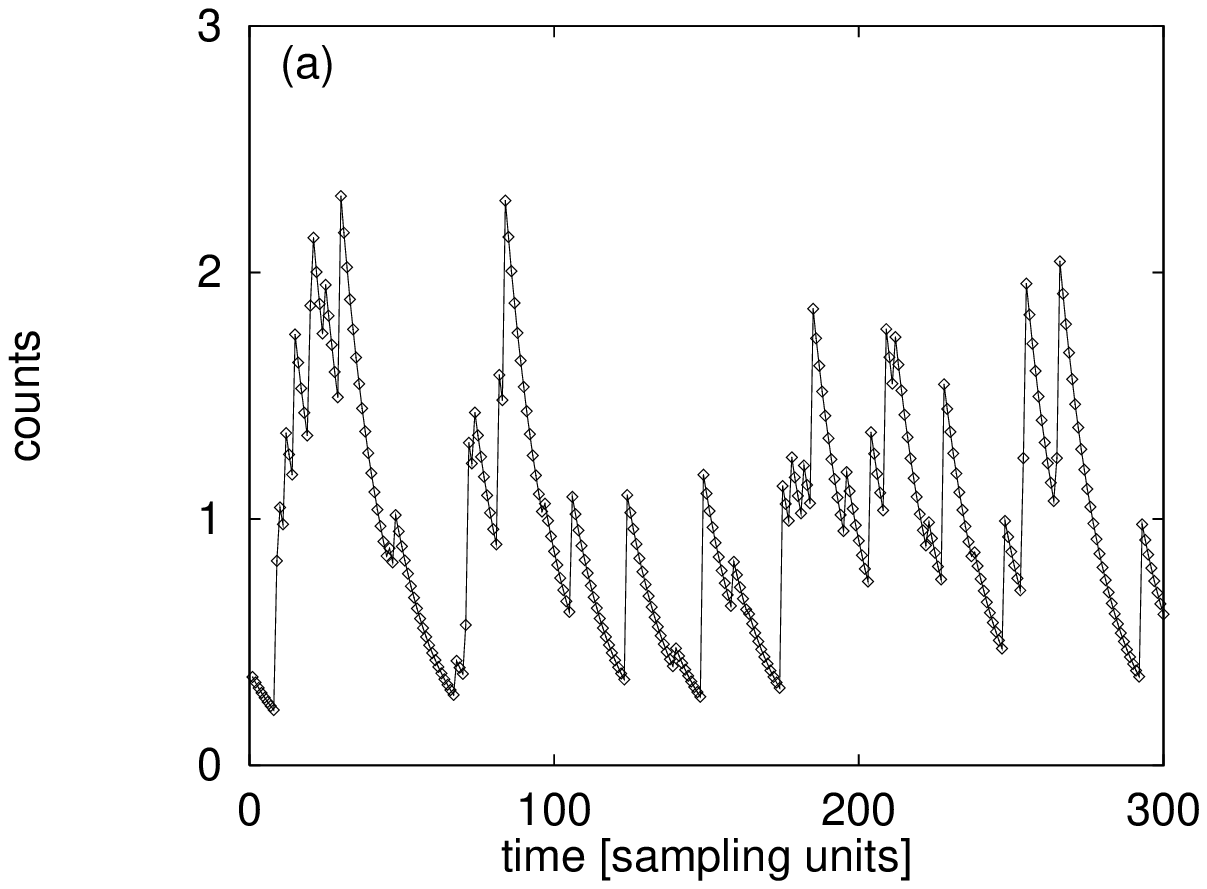}}
\parbox{\textwidth}{
\epsfxsize=0.5\textwidth \epsfbox{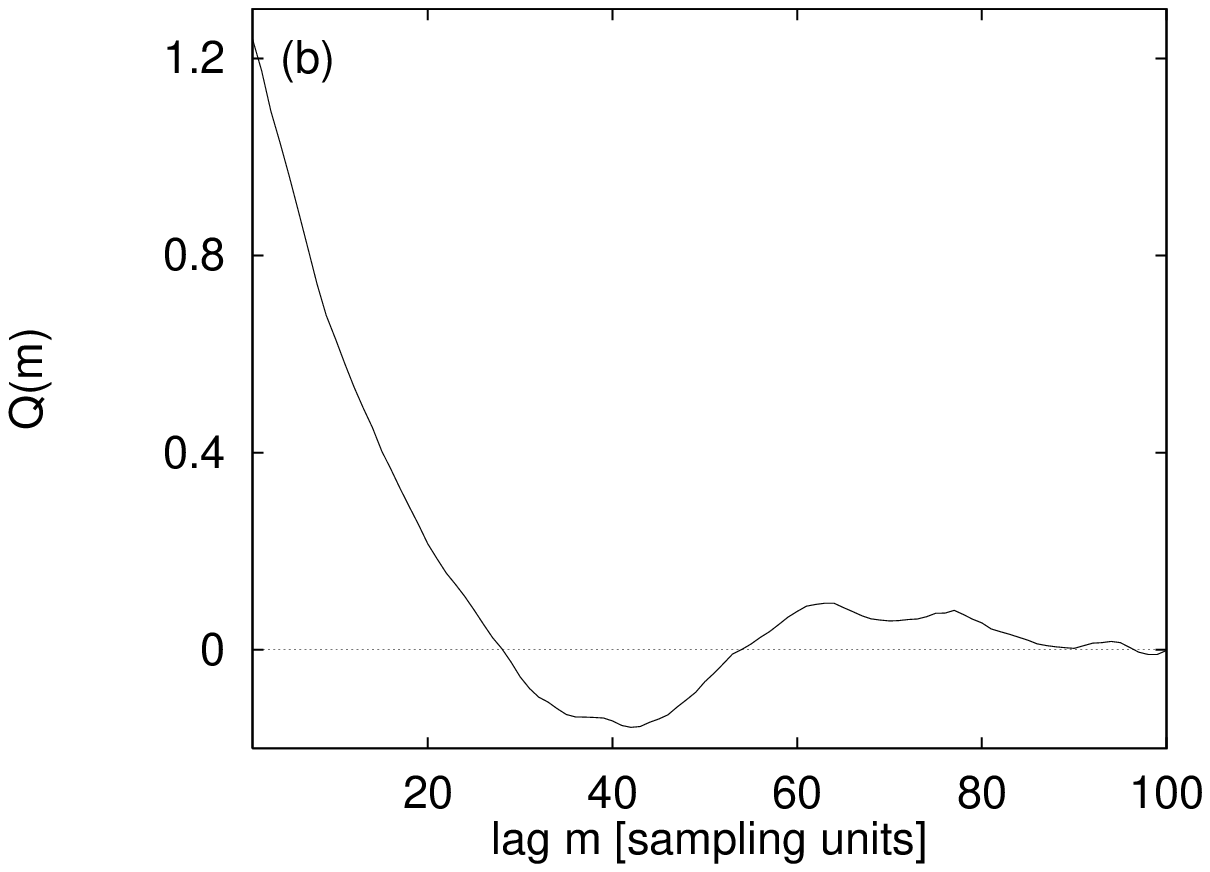}}
\begin{center}
\parbox{\textwidth}{
\epsfxsize=0.5\textwidth \epsfbox{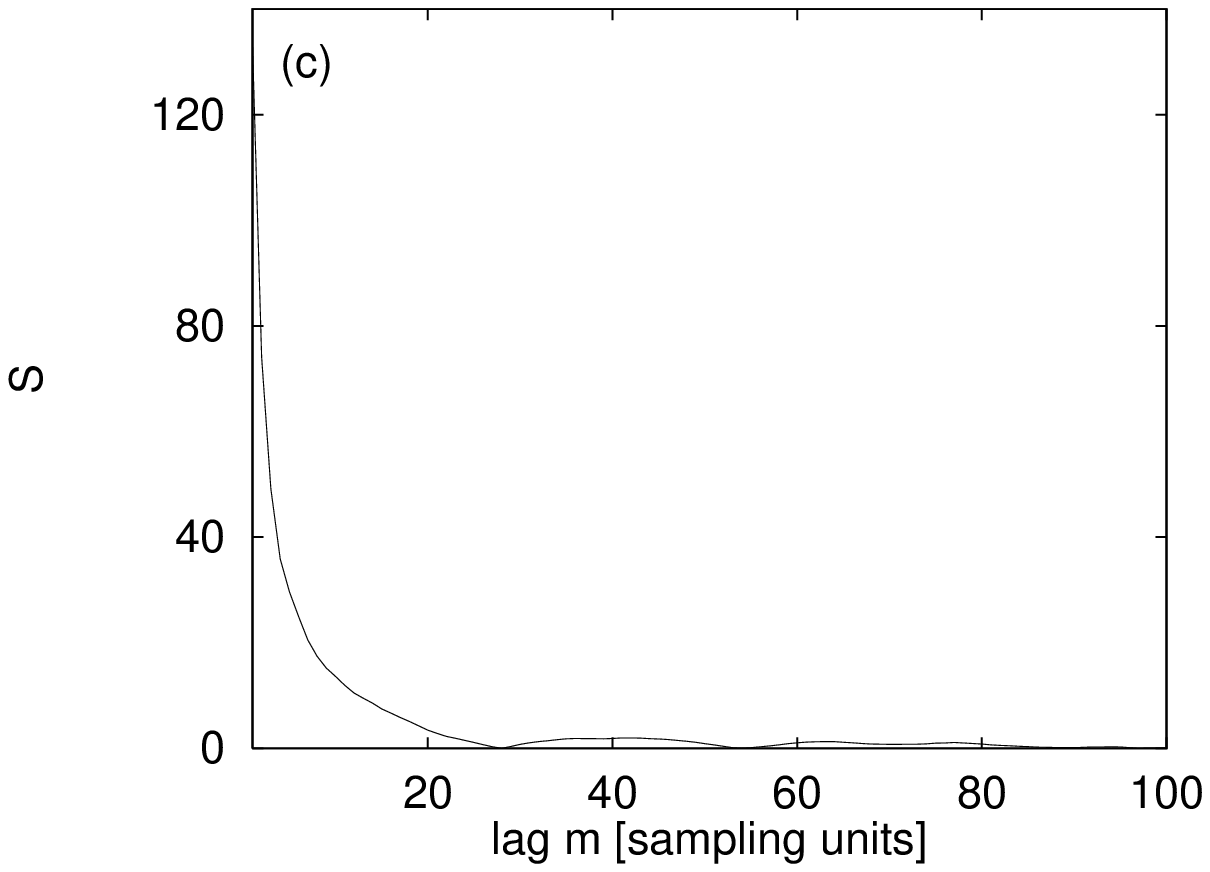}}
\end{center}
\caption[]{\label{fig_q_true}}
\end{center}
\end{figure}

\begin{figure}
\parbox{\textwidth}{
\epsfxsize=0.5\textwidth \epsfbox{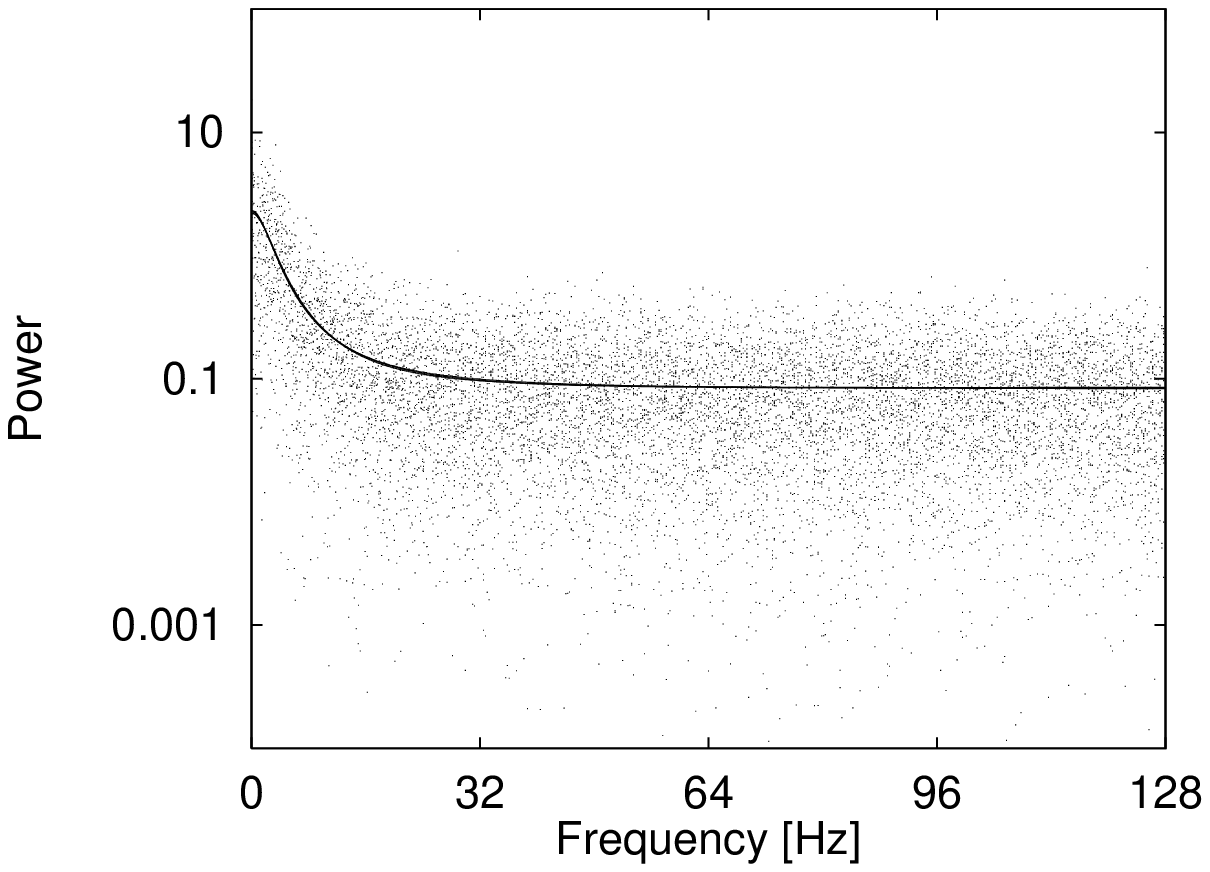}}
\parbox{\textwidth}{
\epsfxsize=0.5\textwidth \epsfbox{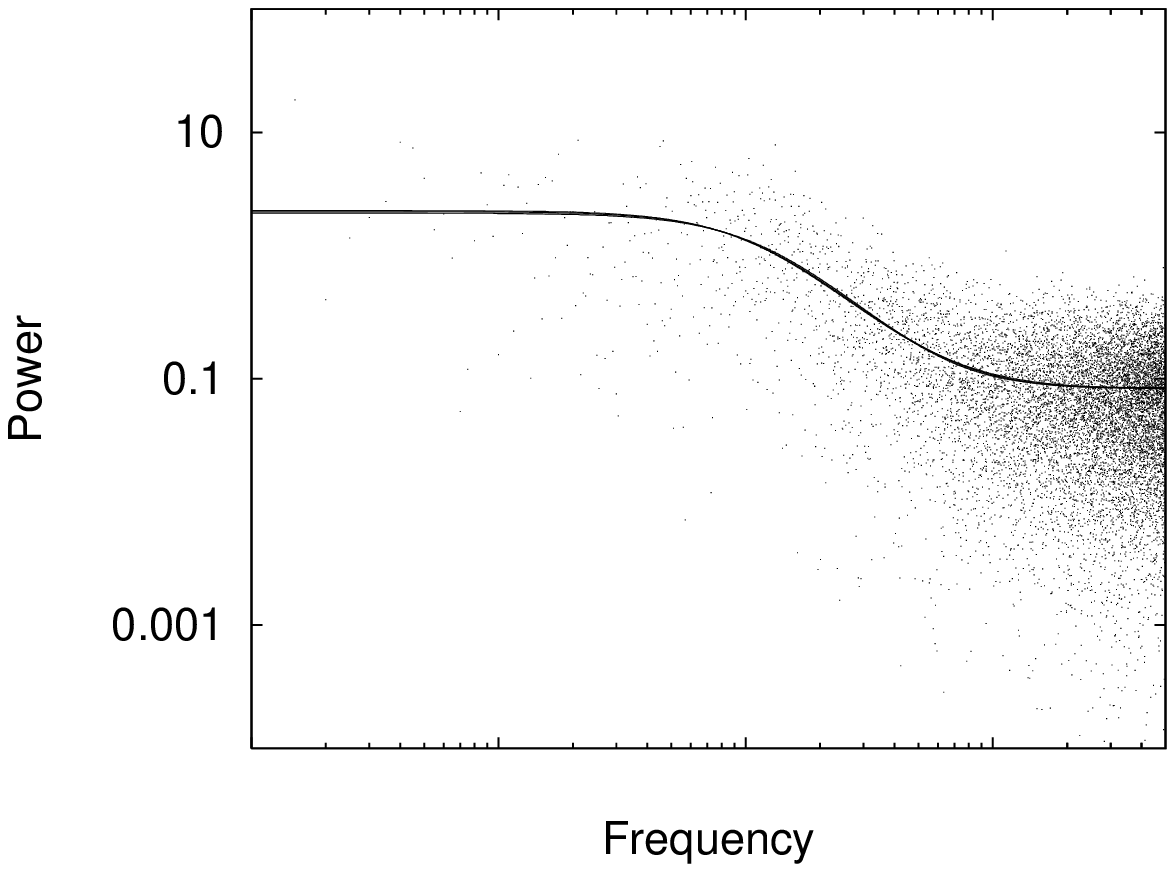}}
\caption[]{\label{fig_spec_lssm}}
\end{figure}

\begin{figure}
\parbox{\textwidth}{
\epsfxsize=0.45\textwidth \epsfbox{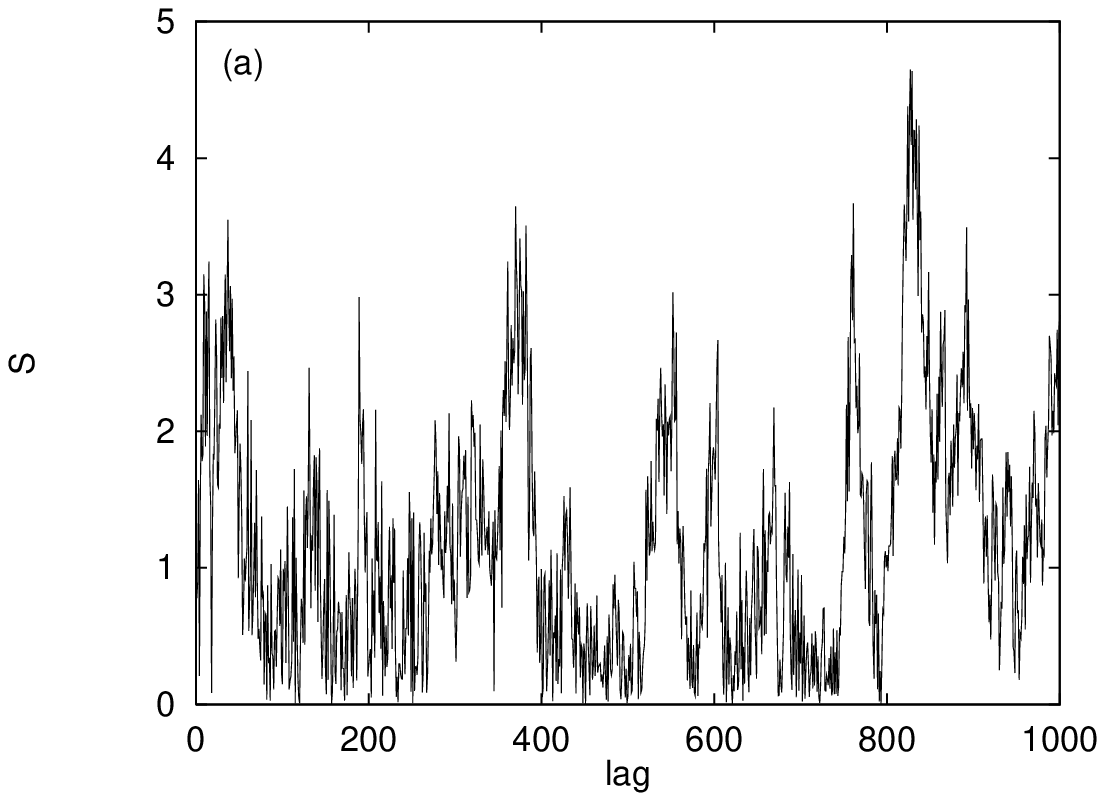}}
\parbox{\textwidth}{
\epsfxsize=0.45\textwidth \epsfbox{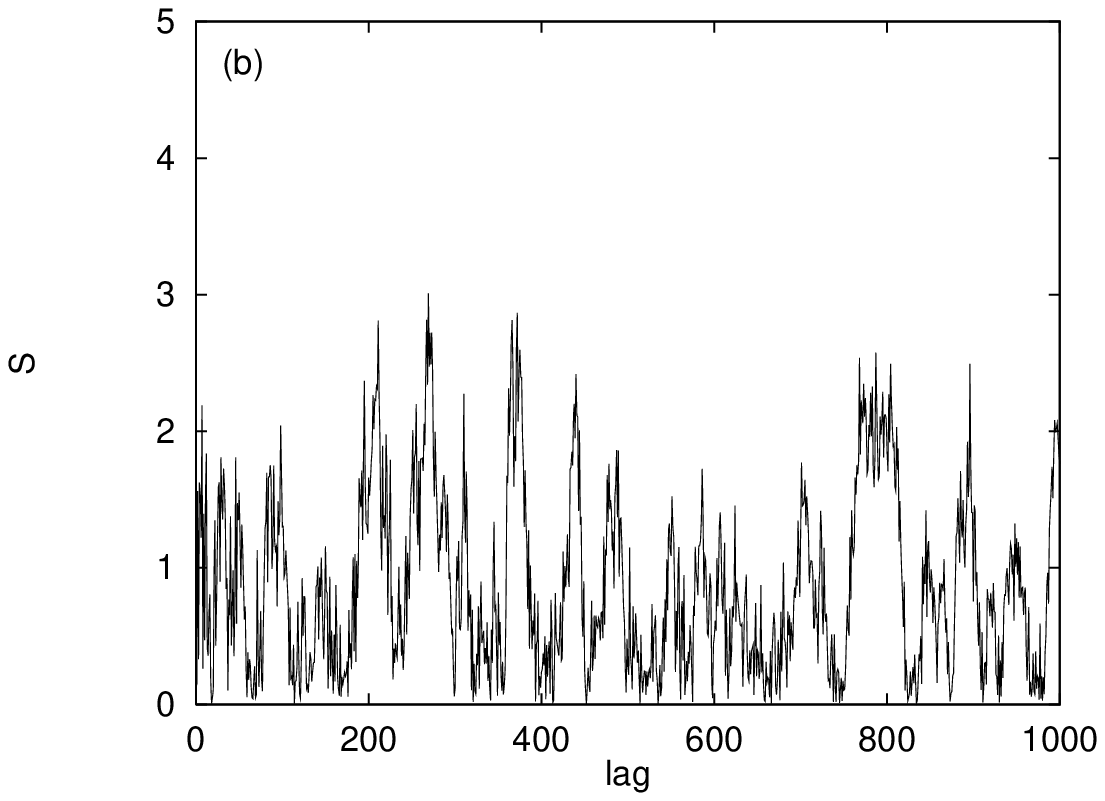}}
\parbox{\textwidth}{
\epsfxsize=0.45\textwidth \epsfbox{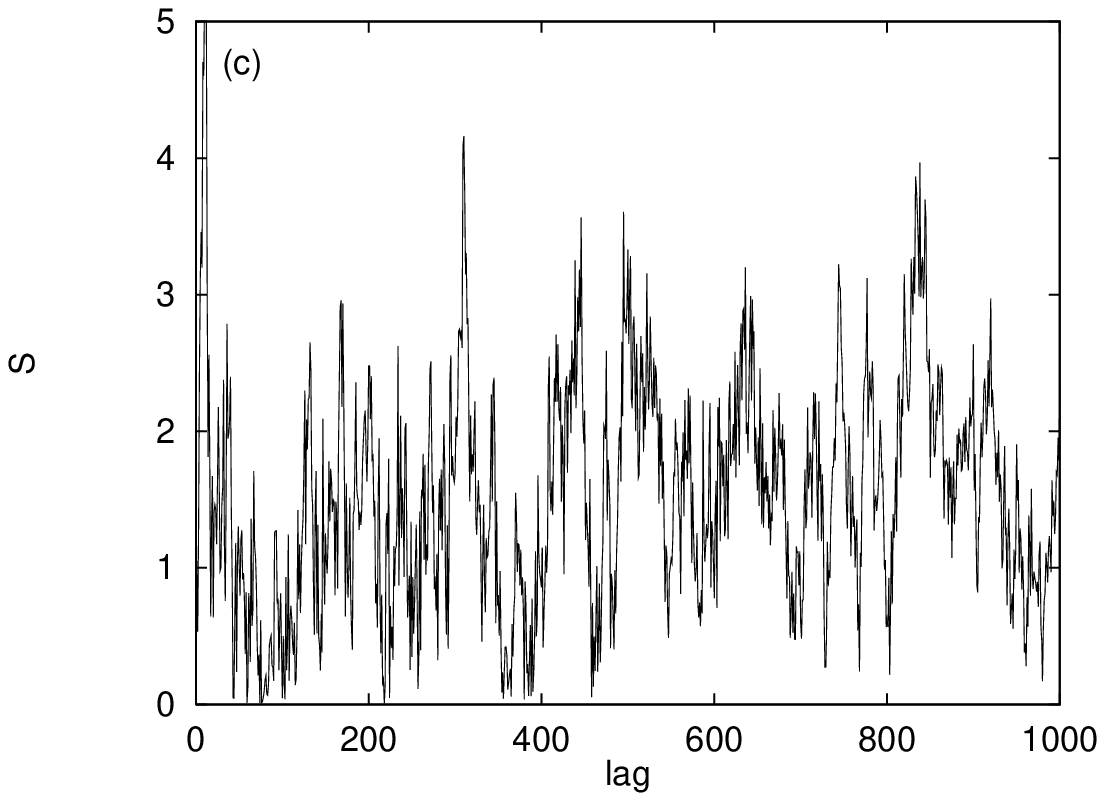}}
\parbox{\textwidth}{
\epsfxsize=0.45\textwidth \epsfbox{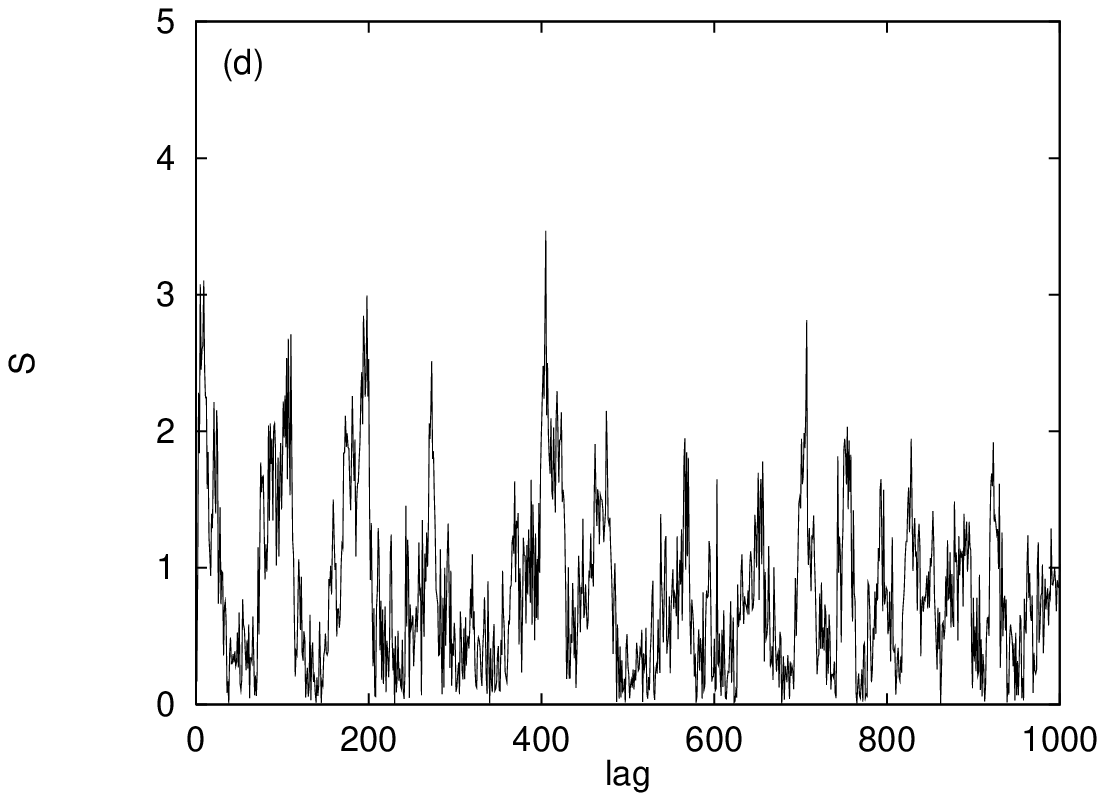}}
\caption[]{\label{fig_q_roh}}
\end{figure}

\begin{figure}
\parbox{\textwidth}{
\epsfxsize=0.45\textwidth \epsfbox{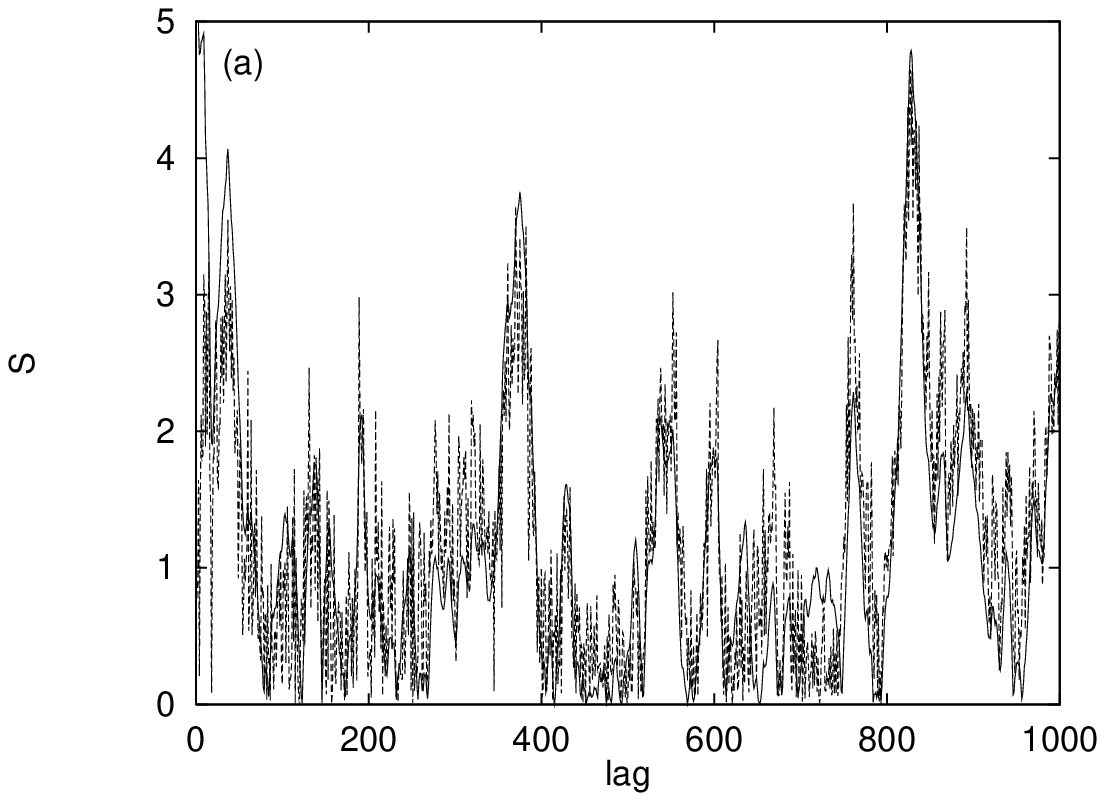}}
\parbox{\textwidth}{
\epsfxsize=0.45\textwidth \epsfbox{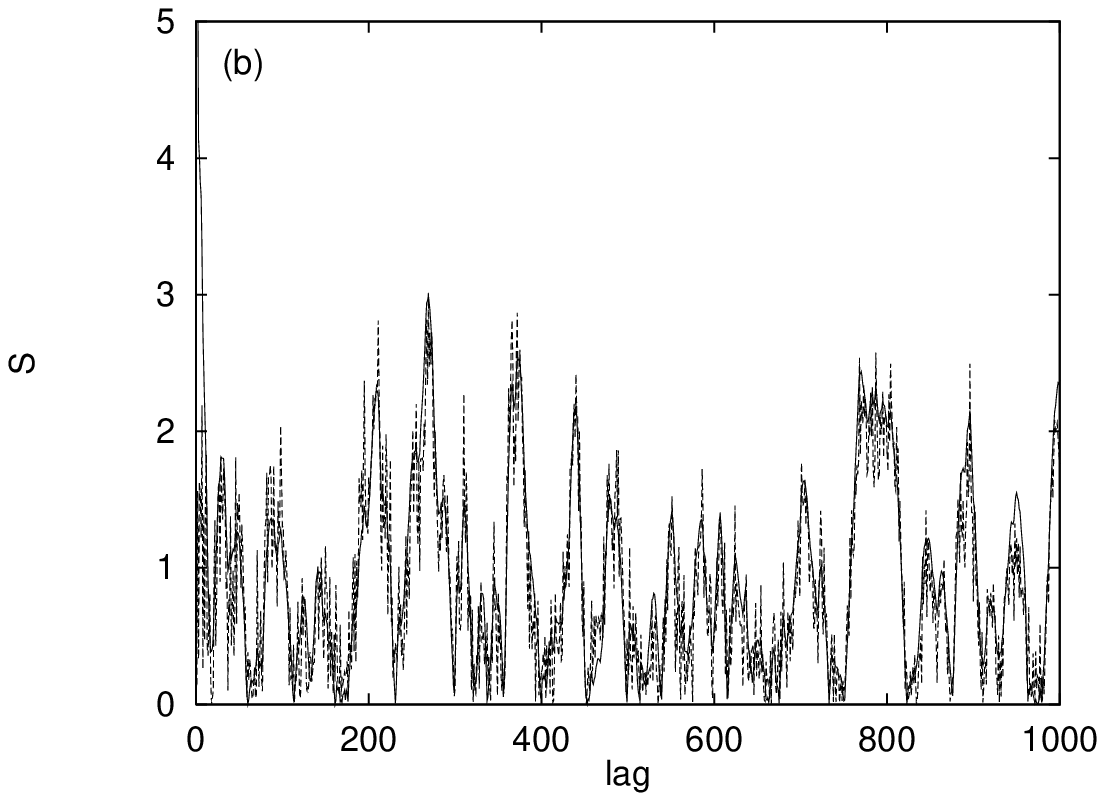}}
\parbox{\textwidth}{
\epsfxsize=0.45\textwidth \epsfbox{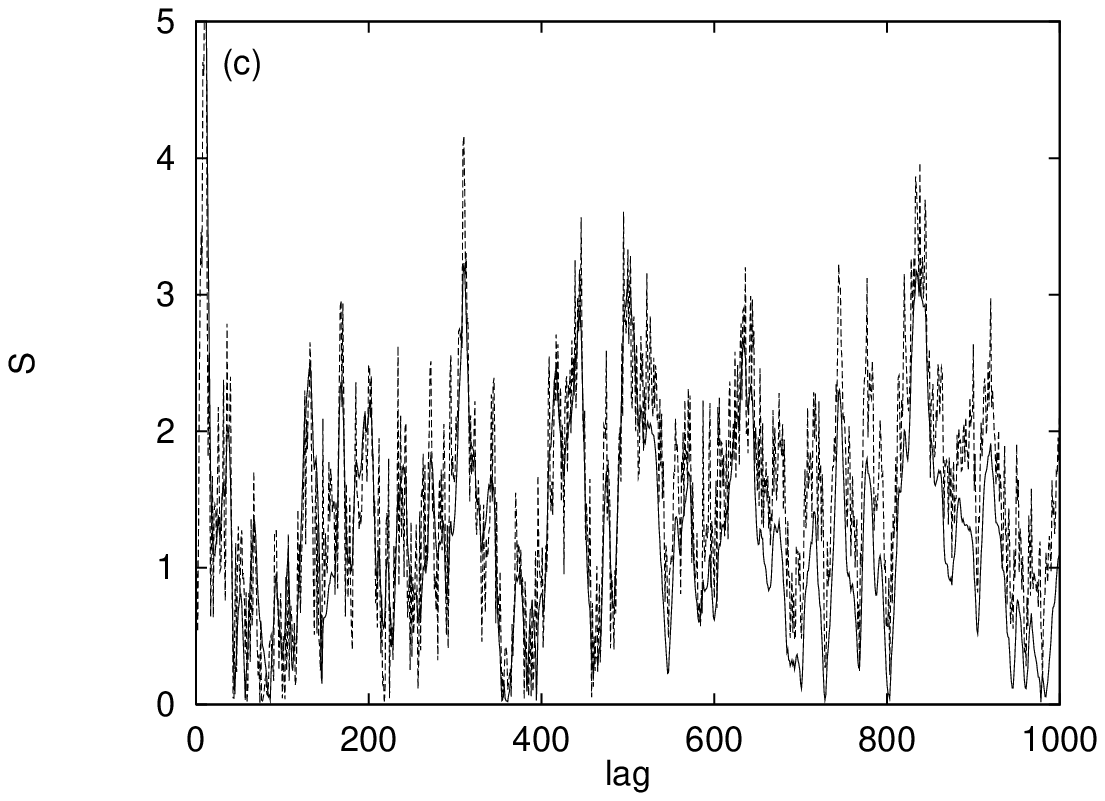}}
\parbox{\textwidth}{
\epsfxsize=0.45\textwidth \epsfbox{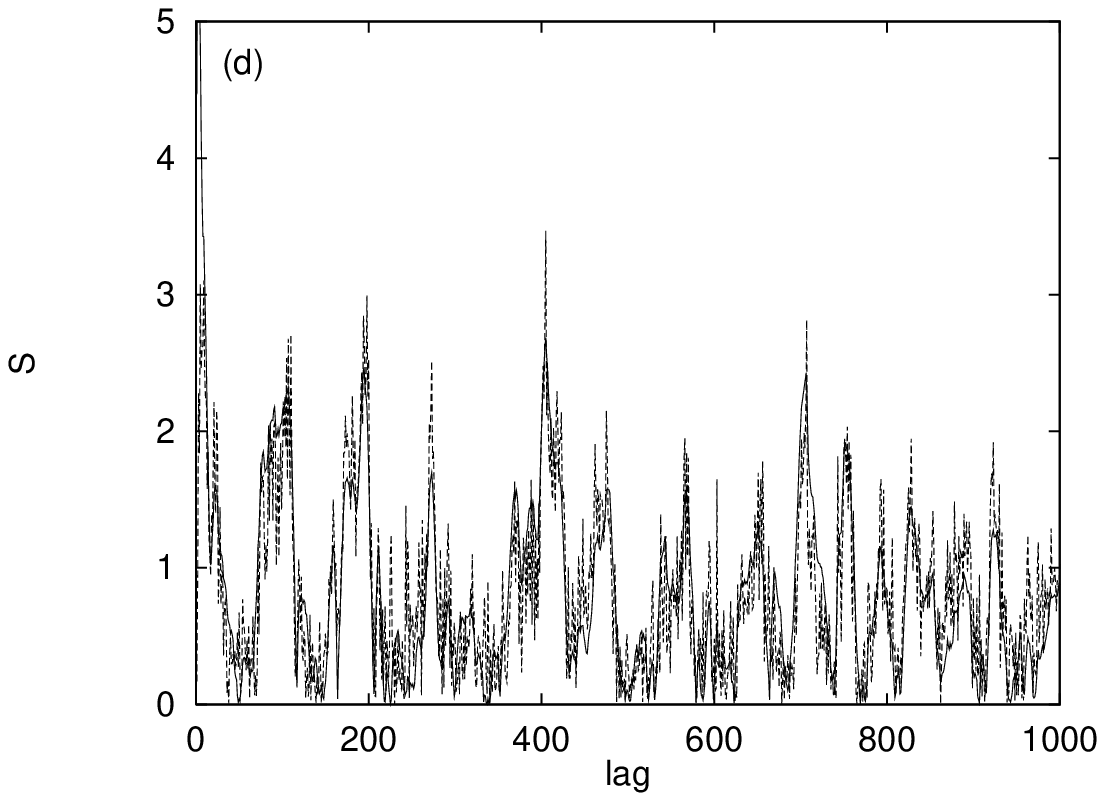}}
\caption[]{\label{fig_q_smooth1}}
\end{figure}

\begin{figure}
\parbox{\textwidth}{
\epsfxsize=0.45\textwidth \epsfbox{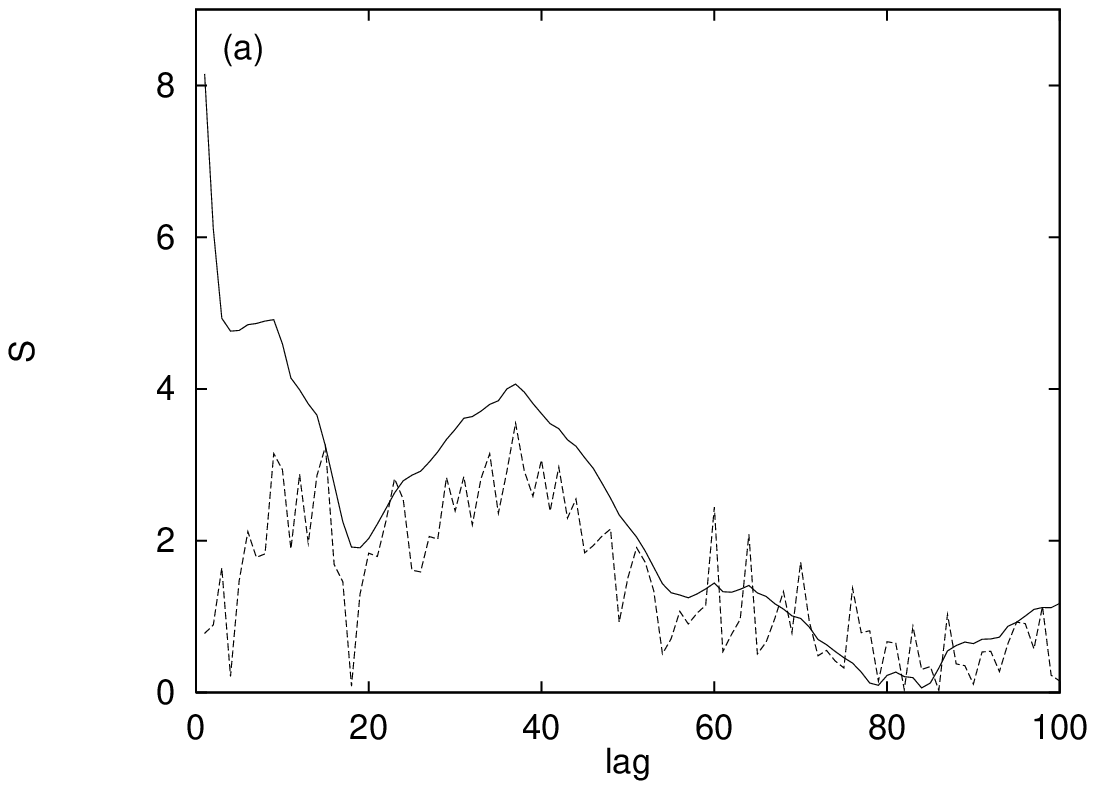}}
\parbox{\textwidth}{
\epsfxsize=0.45\textwidth \epsfbox{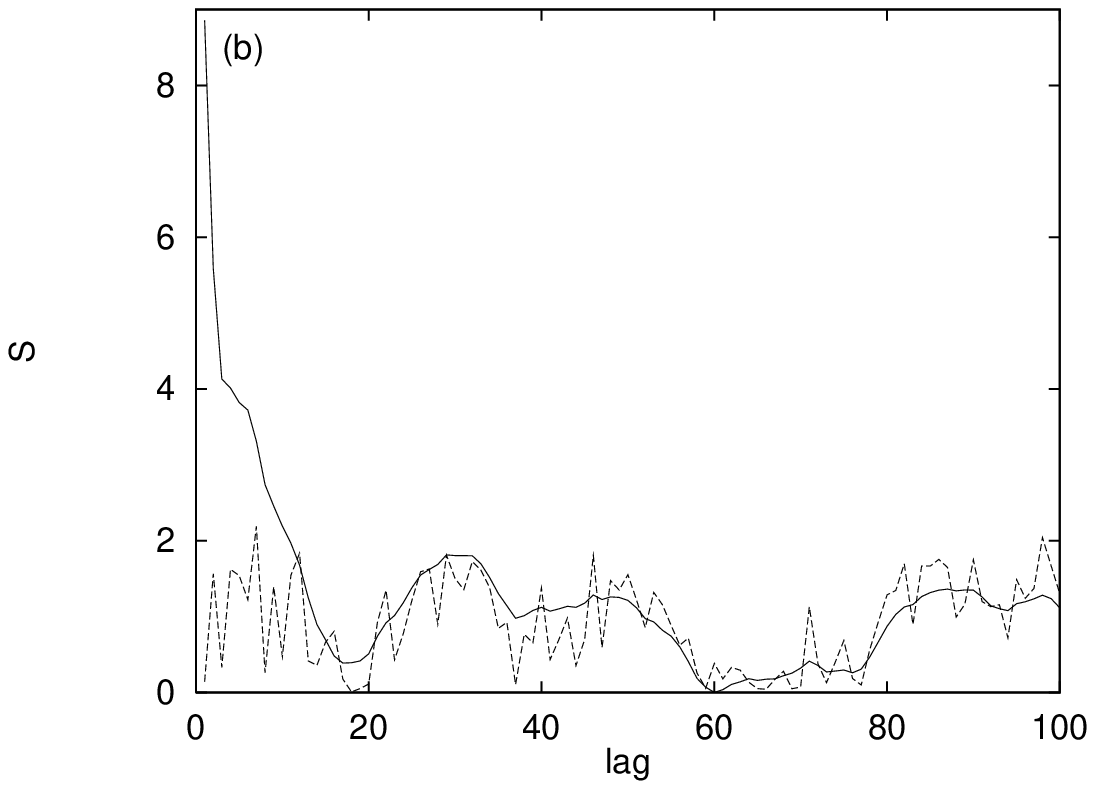}}
\parbox{\textwidth}{
\epsfxsize=0.45\textwidth \epsfbox{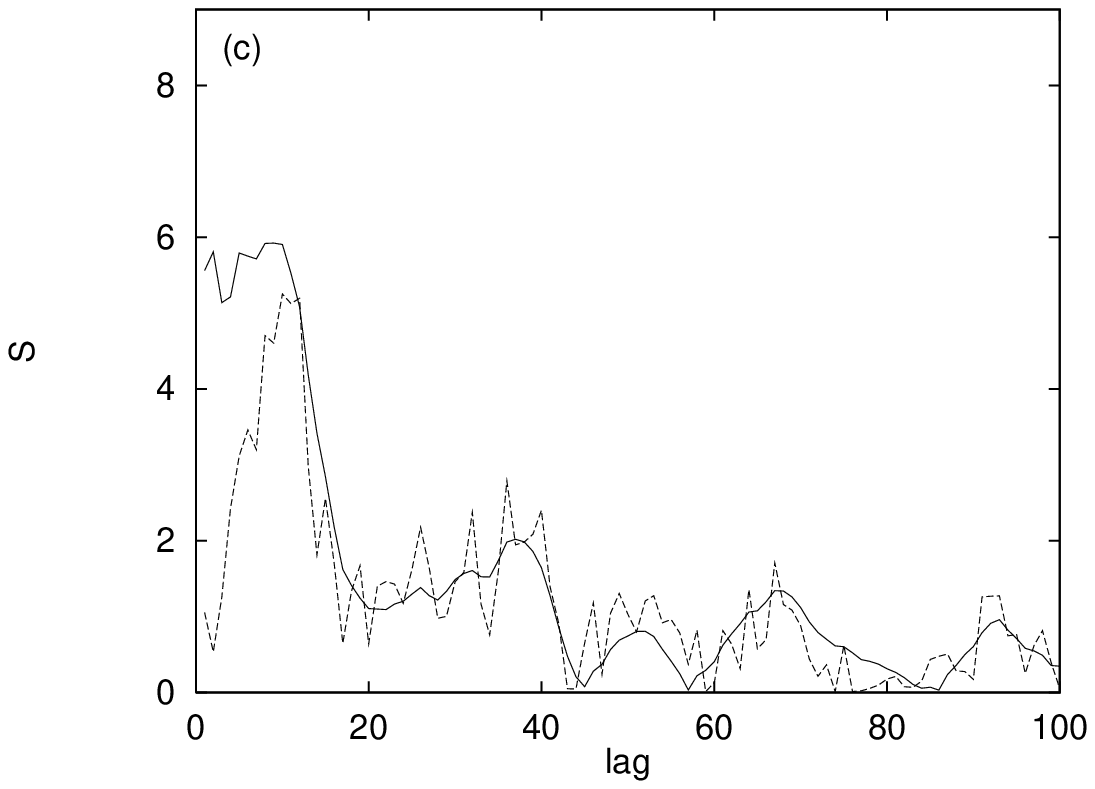}}
\parbox{\textwidth}{
\epsfxsize=0.45\textwidth \epsfbox{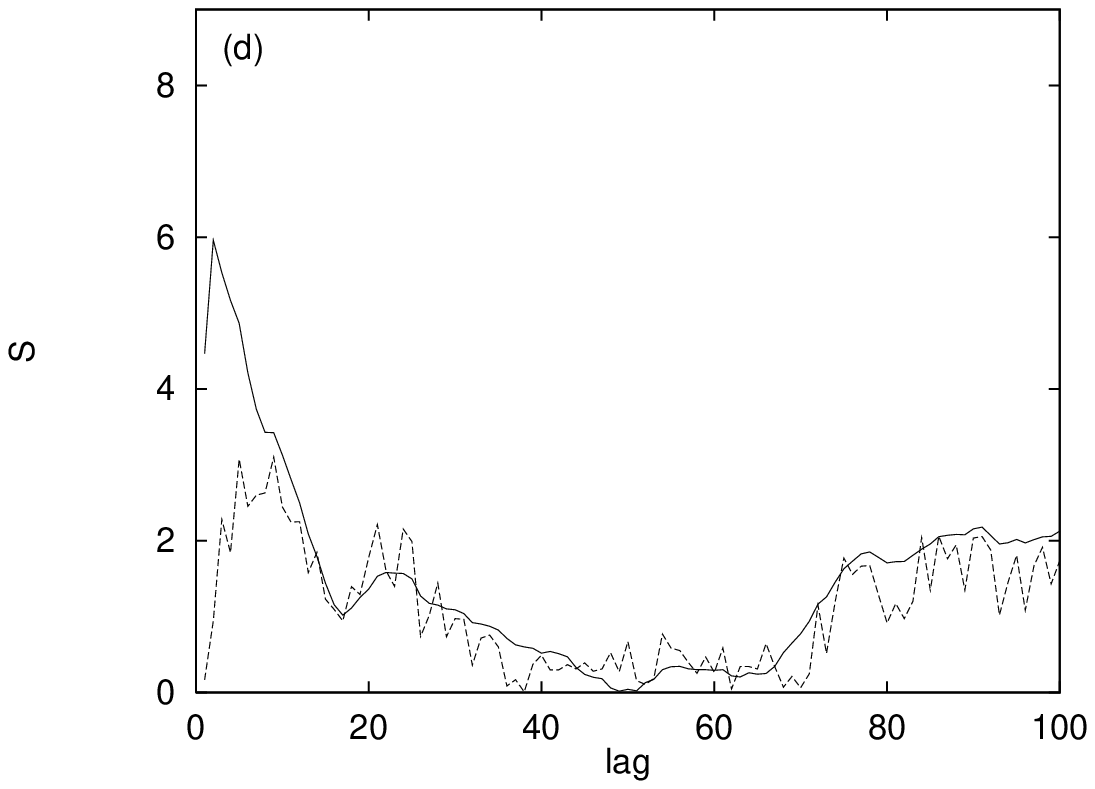}}
\caption[]{\label{fig_q_smooth2}}
\end{figure}

\begin{figure}
\parbox{\textwidth}{
\epsfxsize=0.5\textwidth \epsfbox{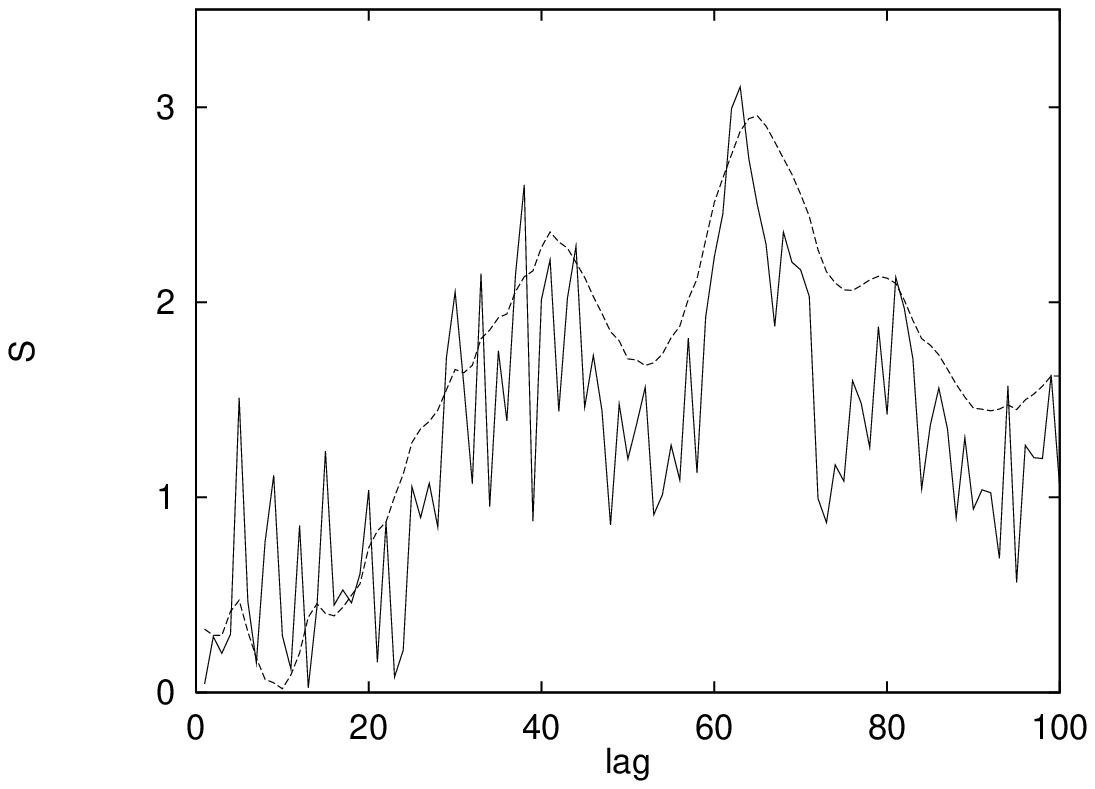}}
\caption[]{\label{fig_artis}}
\end{figure}

\begin{figure}
\parbox{\textwidth}{
\epsfxsize=0.5\textwidth \epsfbox{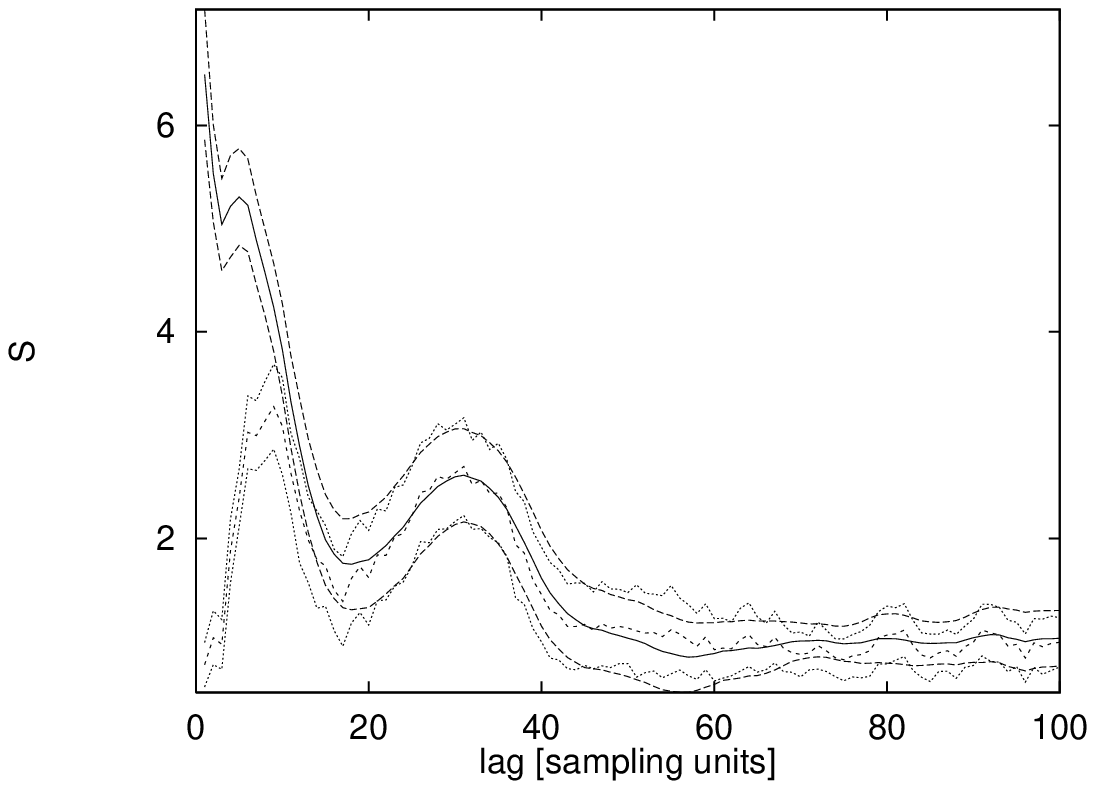}}
\caption[]{\label{fig_q_all}}
\end{figure}

\begin{figure}
\parbox{\textwidth}{
\epsfxsize=0.5\textwidth \epsfbox{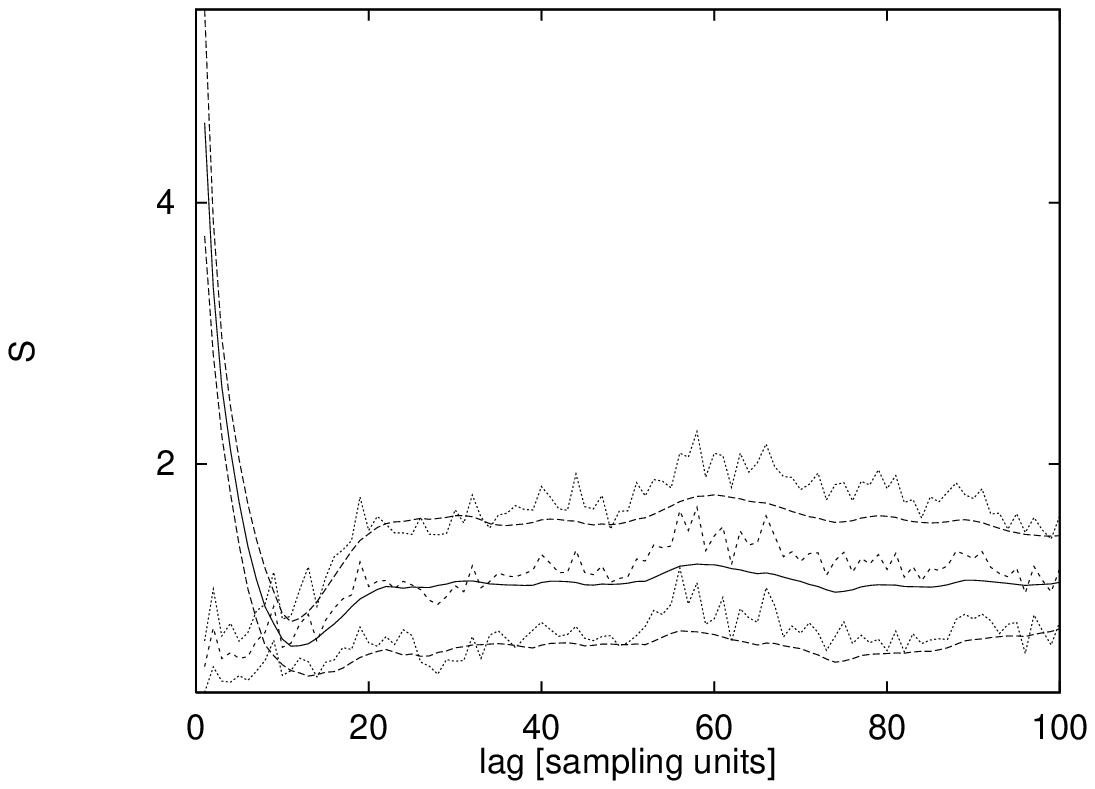}}
\caption[]{\label{fig_q_all_low}}
\end{figure}

\begin{itemize}

\item [Fig.~\ref{fig_data}]
A 3 s segment of the intermediate state time series.

\item [Fig.~\ref{fig_q_true}]  Analysis of a simulated shot noise process. (a)
	Segment of a realization of an exponential shot noise process
        with intensity $\rho=0.1$ and decay time $\tau=15$
	sampling units. (b) The $Q(m)$~statistics, Eq.~(\ref{q_statistic}). 
	(c) Significances $S(m)$, Eq.~(\ref{significances}).

\item [Fig.~\ref{fig_spec_lssm}] Periodogram of the data (dots) and spectra
	(solid lines) calculated from the estimated parameters of the state space
	model of dimension one to three in log-linear scale (top) and log-log
	scale (bottom).
        Note that the spectra are virtually indistinguishable.

\item [Fig.~\ref{fig_q_roh}] Significances $S(m)$ of the $Q(m)$~statistics for
	 lags up to 1000. (a)  First segment of the intermediate state data set. 
	(b--d) Results for the second to the fourth segment.

\item [Fig.~\ref{fig_q_smooth1}] Significances $S(m)$
	analogous to Fig.~\ref{fig_q_roh}. Dashed lines: Results for
	the raw data. Solid lines: Results for data after Kalman-filtering.

\item [Fig.~\ref{fig_q_smooth2}] Significances $S(m)$ of the $Q(m)$~statistics 
	for lags up to 100. Dashed lines: Results for the raw data. Solid 
	lines: Results for data after Kalman-filtering.

\item [Fig.~\ref{fig_artis}] Results from a simulation study using the
	fitted LSSM. The significances $S(m)$ of the $Q(m)$~statistics are 
	calculated for the
	raw data (solid line) and the data after Kalman-filtering (dashed line).
 
\item [Fig.~\ref{fig_q_all}] Significances $S(m)$ of the $Q(m)$~statistics 
	and 2$\sigma$ confidence regions
	calculated from the 32 segments of length 20,000 of intermediate
	state time series. Dashed line: Raw data.  
	Solid line: Kalman-filtered data.

\item [Fig.~\ref{fig_q_all_low}] Significances $S(m)$ of the $Q(m)$~statistics 
	and twice the standard error 
	calculated from the 19 segments of length 20,000 of the low state
	time series. Dashed line: Raw data.  
	Solid line: Kalman-filtered data.


\end{itemize}

\clearpage

\clearpage

\clearpage

\clearpage

\clearpage

\clearpage

\end{document}